\documentclass[12pt]{article}
\usepackage{epsfig}
\usepackage{amsmath}
\usepackage{hhline}
\usepackage{amssymb}
\usepackage{times}
\usepackage{cite}

\newlength{\dinwidth}
\newlength{\dinmargin}
\newcommand{\etb}{\mbox{$E_{t,B}~$}}
\newcommand{\etbx}{\mbox{$E_{t,B}$}}
\newcommand{\deltab}{\mbox{$\Delta_B~$}}
\newcommand{\deltabx}{\mbox{$\Delta_B$}}
\newcommand{\nb}{\mbox{$n_B~$}}
\newcommand{\nbx}{\mbox{$n_B$}}
\newcommand{\etjet}{\mbox{$E_{t,Jet}~$}}
\newcommand{\etjetx}{\mbox{$E_{t,Jet}$}}
\newcommand{\sph}{\mbox{${\rm Sph_B}$}~}
\newcommand{\sphx}{\mbox{${\rm Sph_B}$}}

\newcommand{\rrho}{\mbox{$R/\langle \rho \rangle$}}
\newcommand{\Qsq}{\mbox{$Q^2$}}

\newcommand{\GeV}{\mbox{\rm ~GeV~}}
\newcommand{\GeVx}{\mbox{\rm ~GeV}}

\newcommand{\GeVsq}{\mbox{${\rm ~GeV}^2~$}}
\newcommand{\GeVsqx}{\mbox{${\rm ~GeV}^2$}}

\newcommand{\pb}{\mbox{${\rm ~pb~}$}}
\newcommand{\pbx}{\mbox{${\rm ~pb}$}}

\newcommand{\lsim}{\raisebox{-1.5mm}{$\:\stackrel{\textstyle{<}}{\textstyle{\sim}}\:$}}
\newcommand{\gsim}{\raisebox{-0.5mm}{$\stackrel{>}{\scriptstyle{\sim}}$}}

\newcommand{\qprimesq}{\mbox{${Q'}^2~$}}

\newcommand{\qprimesqrec}{\mbox{${Q'}^2_{\hspace{-0.15cm} \rm rec}~$}}
\newcommand{\qprimesqrecx}{\mbox{${Q'}^2_{\hspace{-0.15cm} \rm rec}$}}
\newcommand{\xprime}{\mbox{$x'~$}}

\newcommand{\qprimesqx}{\mbox{${Q'}^2$}}
\newcommand{\xprimex}{\mbox{$x'$}}

\begin{document}

\begin{titlepage}

\noindent
DESY 02-062  \hfill  ISSN 0418-9833 \\
May 2002

\vspace*{3cm}
\begin{center}
  \Large
  {\bf 
Search for QCD Instanton-Induced Processes in Deep-Inelastic Scattering at HERA
 }
  \vspace*{1cm} \\
  {\Large H1 Collaboration} 
\end{center}

\begin{abstract}
\noindent
Signals of QCD instanton-induced processes are searched for
in deep-inelastic scattering (DIS) at the electron-proton collider HERA 
in a kinematic region defined by the Bjorken-scaling variables 
$x > 10^{-3}$, $0.1 < y < 0.6$ and 
photon virtualities $10 \lsim Q^2 < 100$ \GeVsqx.
Several observables characterising hadronic final state 
properties of QCD instanton-induced events are exploited 
to identify a potentially instanton-enriched domain.
While an excess of events with instanton-like topology over the expectation of the standard DIS background is observed it  can not be claimed to be significant given the uncertainty of the simulation.
Upper limits on the cross-section for instanton-induced processes of
between $60$\pb and  $1000$\pb
are set dependent on the kinematic domain considered.
The data do not exclude the cross-section predicted by instanton perturbation theory for small instanton sizes. 
At large instanton sizes a naive extrapolation of instanton perturbation theory yields a cross-section in the range of sensitivity of this study. Such a cross-section is not observed, in agreement with non-perturbative lattice simulations of the QCD vacuum.
\end{abstract}
\vfill
\begin{center}
To be submitted to Eur. Phys. J. C
\end{center}

\end{titlepage}

\begin{flushleft}

C.~Adloff$^{33}$,              
V.~Andreev$^{24}$,             
B.~Andrieu$^{27}$,             
T.~Anthonis$^{4}$,             
A.~Astvatsatourov$^{35}$,      
A.~Babaev$^{23}$,              
J.~B\"ahr$^{35}$,              
P.~Baranov$^{24}$,             
E.~Barrelet$^{28}$,            
W.~Bartel$^{10}$
% ...

\bigskip{\it
% ...
$ ^{4}$ Inter-University Institute for High Energies ULB-VUB, Brussels;
          Universiteit Antwerpen (UIA), Antwerpen; Belgium$^{ c}$ \\
% ...
$ ^{10}$ DESY, Hamburg, Germany \\
% ...
$ ^{23}$ Institute for Theoretical and Experimental Physics,
          Moscow, Russia$^{ l}$ \\
 $ ^{24}$ Lebedev Physical Institute, Moscow, Russia$^{ e}$ \\
% ...
$ ^{27}$ LPNHE, Ecole Polytechnique, IN2P3-CNRS, Palaiseau, France \\
 $ ^{28}$ LPNHE, Universit\'{e}s Paris VI and VII, IN2P3-CNRS,
          Paris, France \\
% ...
$ ^{33}$ Fachbereich Physik, Bergische Universit\"at Gesamthochschule
          Wuppertal, Wuppertal, Germany \\
% ...
$ ^{35}$ DESY, Zeuthen, Germany \\
% ...
$ ^c$ Supported by FNRS-FWO-Vlaanderen, IISN-IIKW and IWT \\
% ...
$ ^e$ Supported by the Deutsche Forschungsgemeinschaft \\
% ...
$ ^l$ Partially Supported by Russian Foundation
      for Basic Research, grant    no. 00-15-96584 \\
}
\end{flushleft}

\newpage

\section{Introduction}
The Standard Model of particle physics is known to contain certain processes 
which violate the conservation of baryon and lepton number ($B + L$) 
in the case of electroweak interactions 
and chirality in the case of strong interactions \cite{inst:thooft}. 
Such anomalous processes cannot be 
described by standard perturbation theory. 
They are induced by instantons \cite{inst:thooft,inst:belavin}.
In the strong sector, described by quantum chromodynamics (QCD),
instantons are non-perturbative fluctuations of the gluon field. 
They represent tunnelling transitions between 
topologically non-equivalent vacua.
Deep-inelastic scattering (DIS) offers a unique 
opportunity~\cite{inst:vladimir} to discover a class of hard processes
induced by small QCD instantons. 
The rate is calculable\footnote{For an exploratory calculation of the
instanton-induced contribution to the gluon-induced part of the
total DIS cross-section at 
large values of the Bjorken scaling variable $x > 0.3$, 
see Ref.~\cite{inst:balitsky1}.}
within ``instanton-perturbation theory''
and is found to be sizeable \cite{inst:moch,inst:rs4,inst:rs-lat}.
Moreover, the instanton-induced final state exhibits a characteristic signature
\cite{inst:vladimir,inst:greenshaw,inst:mcws99,inst:schremppzoom,inst:qcdins}.
Detailed reviews are given in Refs.~\cite{inst:ringberg00,inst:ringberg01}.

An experimental observation of  instanton-induced processes 
would constitute a discovery of a basic and novel non-perturbative
QCD effect at high energies.
The theory and phenomenology for the production of instanton-induced processes
at HERA in positron proton collisions at a centre of mass
energy of $300$~{\rm GeV} has recently been worked out by
Ringwald and Schrempp \cite{inst:vladimir,inst:moch,inst:rs4,inst:rs-lat,inst:greenshaw,inst:mcws99}.
The size of the predicted cross-section is large enough     
to make an experimental observation possible.
The expected signal rate is, however, still small compared to that from the
standard DIS process. 
The suppression of the standard DIS background
is therefore the key issue in this analysis.
QCD instanton-induced processes can be discriminated from standard DIS by their 
characteristic hadronic final state signature, 
consisting of a large number of hadrons at high transverse energy
emerging from a ``fire-ball''-like topology in the instanton rest system 
\cite{inst:vladimir,inst:greenshaw,inst:mcws99}.
Derived from simulations studies characteristic observables are exploited to
identify a phase space region where a difference between
data and the standard DIS simulations would indicate a 
contribution from instanton-induced processes.

Upper cross-section limits on instanton-induced processes 
based on standard hadronic final state observables
measured with the H1 detector
have already been derived in Refs.~\cite{h1:k0,h1:mult,carlikuhlen}. 
Here, we present for the first time a dedicated experimental search
for instanton-induced processes in high energy particle collisions.

\section{Phenomenology of QCD Instanton-Induced Processes in DIS}
\label{sec:qcdtheo}
\begin{figure}[h]
   \centering
\hspace{1.0cm}
\begin{tabular}{ll}
\mbox{
 \epsfig{file=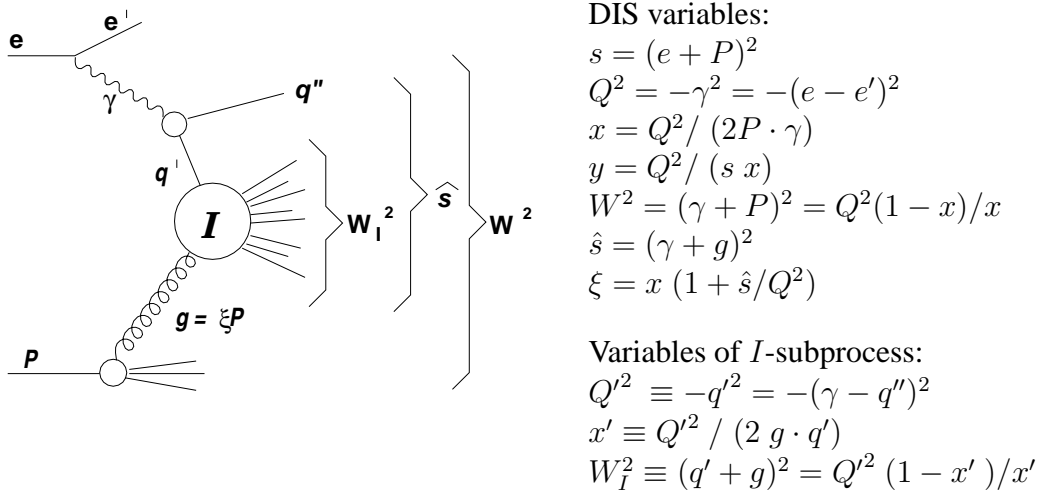,width=7.cm}
}
&
\begin{tabular}{l}
 \vspace{-4.2cm} \\
 DIS variables: \\
   $s=(e+P)^2$\\
   $\Qsq = - \gamma^2 = -(e-e')^2$ \\
   $ x = \Qsq / \; (2 P \cdot \gamma) $ \\
   $ y = \Qsq / \; (s \; x) $ \\
   $ W^2 =(\gamma+P)^2 = \Qsq (1 - x)/x$ \\
   $ \hat{s} = (\gamma+g)^2$ \\
   $ \xi = x \;(1+\hat{s}/Q^2)$ \\ \\
 \vspace{-0.6cm} \\
 Variables of $I$-subprocess: \\
 $\qprimesq  \equiv - {q'}^2 = - (\gamma-q'')^2 $ \\
 $x' \equiv \qprimesq / \;(2 \; g \cdot q' ) $ \\
 $W_I^2  \equiv (q'+g)^2 = \qprimesq ( 1 - \xprime )/ \xprime $\\
\end{tabular}
\end{tabular}
\vspace{0.3cm}
   \caption[Kinematic variables of QCD instanton-induced process
     in deep-inelastic scattering]
     {Kinematic variables of the dominant $I$-induced process in DIS. 
      The virtual photon 
      (4-momentum $\gamma=e-e'$), emitted by the incoming positron $e$,
      fuses with a gluon ($4$-momentum $g$) radiated from the proton
      (4-momentum $P$).
      The gluon carries a fraction $\xi$ of the longitudinal
      proton momentum. The virtual quark entering the 
      instanton subprocess has $4$-momentum $q'$,
      while the outgoing quark ({\it = current quark}) from the 
      photon splitting process 
      has $q^{\prime\prime}$. 
      $W_I$ is the invariant mass of the quark gluon ($q'g$) system
      and $W$ is the invariant mass of the total hadronic system 
      (the $\gamma P$ system).
      $\hat{s}$ is the invariant mass squared of the $\gamma g$ system.}
   \label{kin-var}
\end{figure}

Instanton ($I$) processes in DIS at HERA are discussed within the 
framework of the work of Ringwald and Schrempp \cite{inst:vladimir,inst:moch,inst:rs4,inst:rs-lat,inst:greenshaw,inst:mcws99}.
These processes dominantly occur in a photon gluon ($\gamma g$) 
fusion process as sketched in Fig.~\ref{kin-var}. 
The characteristic $I$-event signatures result from the following basic reaction:
\begin{equation}
\gamma^* + g \stackrel{(I)}{\rightarrow} \sum_{n_f} (q_R + \bar{q}_R) + \, n_g \, g,
\; \; \;  ( I \rightarrow \bar{I}, R \rightarrow L),
\end{equation}
where $g$, $q_R$ ($\bar{q}_R$) denotes gluons, 
right-handed quarks (anti-quarks), $n_f$ is the number of quark flavours 
and $n_g$ is the number of gluons produced.
Right-handed quarks are produced in $I$-induced processes,
left-handed quarks are produced in anti-instanton $(\bar{I}$) processes. 
The final state induced by instantons or anti-instantons
can only be distinguished by the chirality of the quarks.
Experimental signatures sensitive to instanton-induced chirality violation
are not exploited in this analysis. Both 
$I$-processes and  $\bar{I}$-processes enter in the calculation 
of the total cross-section.

As shown in Fig.~\ref{kin-var}, 
a photon splits into a quark anti-quark pair
in the background of an instanton or an anti-instanton field.
The so-called $I$-subprocess $q' + g  \stackrel{(I,\bar{I})}{\rightarrow} X$
is produced by
the quark or the anti-quark fusing with a gluon $g$ from the proton. 
The respective partonic final state includes $2 \, n_f - 1$ light quarks and anti-quarks. 
Therefore, together with the current quark ($q^{\prime\prime}$),
in every $I$-event, quark anti-quark pairs of each
of the $n_f(=3)$ (light) flavours are simultaneously 
produced\footnote{
In principle, also heavy flavours contribute whenever very small
instantons are probed. In general, however, the quarks must appear
approximately massless on the scale of the dominant effective
$I$-size $\rho_{\rm eff}(\qprimesqx,\xprimex)$, 
i.e. $\rho_{\rm eff} \, m_q \ll 1$, where $m_q$ is the quark mass.
In the HERA kinematic region, the rate is dominated by
$\rho_{\rm eff} \approx 0.35 ~{\rm fm}$ such that only up, down and strange
quarks appear massless ($n_f=3$).
The contribution of charm and bottom quarks to the
cross-section is likely to be small.
It was checked that the predicted final state signature does not
change significantly if heavy quarks are included in the simulation.}.
In addition, a mean number of 
$\langle n_g \rangle  \sim {\cal O}(1/\alpha_s)  \sim  3$ gluons is
expected to be emitted in the $I$-subprocess.

The quarks and gluons emerging from the $I$-subprocess are isotropically 
distributed in the $I$-rest system defined by $\vec{q'} + \vec{g} = 0$.  
One expects therefore a
pseudo-rapidity\footnote{The pseudo-rapidity of a 
particle is defined as $\eta \equiv - \ln  \tan ( \theta / 2) $, where $\theta$ is 
the polar angle with respect to the proton direction defining the
$+z$-axis.}  ($\eta$)
region with a width of typically $2$ units in  $\eta$.
This region is
densely populated with particles of relatively high transverse momentum
which are homogeneously distributed in azimuth in the $I$-rest frame. 
Apart from this pseudo-rapidity band, 
the hadronic final state exhibits a current jet emerging from
the outgoing current quark $q''$.
The large number of partons emitted in the $I$-process leads
to a high multiplicity of charged and neutral particles in every event.

The actual number of produced hadrons and their energies
crucially depends on the centre of mass energy $W_I$ 
available in the $I$-system, which in turn can be written (see Fig.~\ref{kin-var}) 
in terms of the variables  \qprimesq and \xprime
describing the kinematics of the $I$-subprocess. 
These variables are defined in analogy 
to the Bjorken scaling variables $x$ and $Q^2$.
A knowledge of the distributions of these variables
is indispensable for the correct prediction of the hadronic final state.
These distributions can be calculated within $I$-perturbation 
theory \cite{inst:moch,inst:rs4}
for large  \qprimesq and \xprimex.

The total $I$-production cross-section at HERA,
$\sigma^{(I)}_{\rm HERA}$, is essentially determined by 
the cross-section of the $I$-subprocess $q' + g  \stackrel{(I)}{\rightarrow} X$ denoted by
$\sigma_{q'g}^{(I)}$. The latter can be calculated by 
integrating\footnote{For simplicity, the additional integration
over the relative $I \overline{I}$ colour orientation 
has already been performed in equation (\ref{eq:IXsec}).
Both instanton and anti-instanton
degrees of freedom enter in the cross-section formula, since
it is obtained from taking the modulus squared of the amplitude 
depicted in Fig.~\ref{kin-var}.
The complete formula and more details can, for instance, 
be found in Ref.~\cite{inst:rs4} which contains as well the explicit
physical interpretation of the variables $\rho$ and $R$.} 
over the $I$ ($\bar{I}$)-size  $\rho$ $(\bar{\rho})$ and the 
$I\overline{I}$ distance $4$-vector $R_\mu$:
\begin{equation}
  \label{eq:IXsec}
  \sigma_{q'g}^{(I)}(\xprimex, \qprimesqx) = \int d^4R\; e^{i (g + q') \cdot R}
\int_0^{\infty} \! \! \! \! d\rho \int_0^{\infty} \! \! \! \! d\bar{\rho} \;
e^{- (\rho + \bar{\rho}) Q'} 
D(\rho) D(\bar{\rho}) \; \ldots \; 
  e^{-\frac{4\pi}{\alpha_s(\mu_r)} \Omega(R^2/\rho \bar{\rho}, \bar{\rho}/ \rho)}
\end{equation}
where several parts of the integrand have been omitted.
$D(\rho)$ ($ D(\bar{\rho})$) is the $I$-size ($\bar{I}$-size) distribution
that is calculable within $I$-perturbation theory \cite{inst:thooft}
for $\alpha_s(\mu_r) \ln{(\rho \, \mu_r)} \ll 1$ with
$\alpha_s(\mu_r)$ being the strong coupling taken at the renormalisation
scale $\mu_r$ and $N_C = 3$ for QCD~\cite{inst:thooft,inst:bernard,inst:morris}:
\begin{equation}
D(\rho) = d {\left [ \frac{2 \pi}{\alpha_s(\mu_r)} \right ]}^{2 N_c} \;
e^{- \; \frac{2 \pi}{\alpha_s(\mu_r)}} \; \frac{(\mu_r \rho)^{\frac{11}{3} N_C - \frac{2}{3} n_f 
+ {\cal O}(\alpha_s)}}{\rho^5},
\label{eq:drho}
\end{equation}
where $d$ is a known scheme dependent constant.

The function $\Omega(R^2/\rho \bar{\rho}, \bar{\rho}/ \rho)$, with
$ -1 < \Omega(R^2/\rho \bar{\rho}, \bar{\rho}/ \rho) \lsim 0$, describes
the $I \bar{I}$-interaction associated with a resummation of final state
gauge bosons. It is calculable
in $I$-perturbation theory, formally for
$R^2/\rho \bar{\rho} \gg 1$, and may attenuate 
to some extent the exponent $- 2 \pi /\alpha_s$ 
of the exponential in equation (\ref{eq:drho})
that is typical for tunnelling transitions.
For a general $SU(N_C)$ gauge theory with coupling $\alpha$, 
equations (\ref{eq:IXsec}) and (\ref{eq:drho}) 
give the qualitative behaviour for the $I$-cross-section:
\begin{equation}
\sigma_{q'g}^{(I)}  \sim 
{\left [\frac{2 \pi}{\alpha} \right ]}^{4 N_C} 
e^{- \frac{4 \pi}{\alpha} (1 + \Omega)}.
\label{eq:dsigma}
\end{equation}
Thus, in the absence of final state gauge boson effects
(i.e.~$\Omega =0$), with typical values of $\alpha_s \approx 0.4$ at HERA
and the weak gauge coupling $\alpha_w \approx 0.033$,
equation (\ref{eq:dsigma}) illustrates the strong suppression
of electroweak instanton effects which is absent in QCD:
\begin{equation}
{\left [ \frac{2 \pi}{\alpha_s} \right ]}^{12} 
e^{- \frac{4 \pi}{\alpha_s}} \approx 5 \; \; \gg \; \; 
{\left [ \frac{2 \pi}{\alpha_w} \right ]}^{8} 
e^{- \frac{4 \pi}{\alpha_w}} \approx 7 \cdot 10^{-148}. 
\end{equation}
In this picture the tiny instanton-induced electroweak
$B + L$ violation will only be observable, if
the final state emission of a huge number of $W$-bosons counteracts the 
exponential suppression \cite{ringwald},
i.e. $(1 + \Omega) \approx 0$.
In QCD, however, final state gluons are expected to
only provide a moderate numerical
correction of the rate. Correspondingly, the predictions of the $I$-induced
rate in QCD depends much less on the resummation of final state gauge
bosons.

According to equation (\ref{eq:drho})
the $I$-size distribution follows a power law:
\begin{equation}
  \label{eq:pIDensity}
  D(\rho) \sim \rho^{6 - 2/3 n_f + {\cal O}(\alpha_s)}
\end{equation} 
and the integral over $\rho$ ($\bar{\rho}$) generally diverges for large $\rho$ ($\bar{\rho}$).
However, in the DIS regime the exponential factor $e^{- (\rho + \bar{\rho}) Q'}$
appearing in equation  (\ref{eq:IXsec})
ensures the convergence of the integral. For large enough \qprimesq
effectively only small size instantons contribute to the cross-section.
Therefore the $I$-cross-section is calculable in DIS \cite{inst:moch}.

The $\rho$ and the $R/\rho$ distributions can be calculated 
using quenched (for $n_f=0$), non-perturbative lattice simulations
of the QCD vacuum.
They will be discussed and compared with perturbative predictions in
section~\ref{sec:limits}.
By  confronting $I$-perturbation theory with these lattice simulations,
limits on the validity of  $I$-perturbation theory have been derived
\cite{inst:rs4,inst:rs-lat,inst:schremppzoom}.
The calculations agree for 
$\rho \lsim 0.35 ~{\rm fm}$ and $R/\rho \; \gsim \, 1.05$,
which can be translated into limits on \xprime and \qprimesqx, i.e. 
$Q'/ \Lambda^{n_f}_{\overline {MS}} \; \gsim \; 30.8$ 
and $\xprime \gsim ~0.35$ \cite{inst:rs4,inst:schremppzoom}
where $\Lambda^{n_f}_{\overline {MS}}$ is the QCD scale
in the $\overline {MS}$ scheme for $n_f$ flavours.

For the region, 
$\qprimesq  >  (30.8 \; \Lambda_{\overline{MS}}^{(3)})^2  = 64$ \GeVsqx,
$\xprime > 0.35$, 
$x > 10^{-3}$ and $0.1 < y < 0.9$, the $I$-cross-section at HERA
has been estimated 
to be $\sigma^{(I)}_{\rm HERA} \approx 126 \, {\rm pb}$ \cite{inst:rs4}.
This result has recently been
updated \cite{inst:qcdins,inst:ringberg00,inst:schremppdis00} by using the 1998 world
average of the strong coupling \cite{pdg98} to
$ \sigma^{(I)}_{\rm HERA} = 89_{-15}^{+18} \; {\rm pb}$. 
The quoted errors for the $I$-induced cross-section  $\sigma^{(I)}_{\rm HERA}$
only contain the uncertainty 
obtained from varying the strong coupling. 
The change in $\Lambda_{\overline{MS}}^{(3)}$ 
leads to a change of the minimal required \qprimesq to 
$ \qprimesqx_{\rm \hspace{-3mm} min} = 113$ \GeVsqx.
This cross-section has been derived for three-flavours,
corresponding to $\Lambda_{\overline{MS}}^{(3)}= 346^{+31}_{-29} \; {\rm MeV}$.
An additional cut, $Q^2 > {Q'}^2_{\rm min}$, is
advocated \cite{inst:moch,inst:schremppzoom,inst:qcdins}
to reduce remaining theoretical uncertainties connected 
with non-planar diagrams.

In this domain the cross-section is 
$\sigma^{(I)}_{\rm HERA} = 29_{-7.5}^{+10} \; {\rm pb}$.
The calculation is based
on a two loop renormalisation group invariant expression of the
$I$-density $D(\rho)$ and thus does not depend much 
on the chosen renormalisation scale.
%
In the kinematic domain in which this pioneering analysis is performed, i.e.
the polar angle of the scattered positron
$\theta_{e} > 156^\circ$, $ 0.1 < y < 0.6$, $x >10^{-3}$
and $10 \lsim Q^2< 100~{\rm GeV}^2$, 
the cross-section calculated with QCDINS\footnote{ 
In this result theoretical uncertainties connected with non-planar diagrams
are not taken into account. However, the observables used in this analysis
to calculate cross-section limits seem to be rather insensitive
to these uncertainties \cite{inst:schremppzoom}.}
is $\sigma^{(I)}_{\rm HERA} = 43 \; {\rm pb}$.

Even though these predictions have not yet reached the same quantitative
level of precision as current standard perturbative QCD calculations, 
the cross-section is large enough to motivate
dedicated searches for $I$-processes at HERA.

\section{The H1 Detector at HERA}\label{detector}
A detailed description of the H1 detector can be found elsewhere~\cite{H1det}.
Here we briefly introduce the detector components most relevant 
for this analysis: the liquid argon (LAr) calorimeter,
the backward lead-fibre calorimeter (SpaCal) 
and the tracking chamber system.

The hadronic energy flow is mainly measured by the
LAr calorimeter~\cite{h1lar} extending over the polar angle range
$4^\circ < \theta <  154^\circ$ with full azimuthal coverage.
It consists of an electromagnetic section ($20-30$ radiation lengths) 
with lead absorber and a hadronic section with steel absorber.
The total depth of both calorimeters 
varies between $4.5$ and $8$ interaction lengths. 
Test beam measurements of the LAr~calorimeter modules show an
energy resolution 
of $\sigma_E/E \approx 0.12/\sqrt{E \;[\GeV]} \oplus 1 \% $
for electromagnetic showers \cite{electrons} and of
$\sigma_{E}/E\approx 0.50/\sqrt{E\;[\GeV]} \oplus 2\%$  
for charged pions~\cite{pions}.

The backward lead-fibre calorimeter SpaCal~\cite{h1SpaCal}
covers the polar angle range $153^\circ < \theta <  177^\circ$. 
In the electromagnetic section, with  a depth of $28$ radiation lengths,  
the position and the energy of electrons are measured. 
The electron energy resolution is 
$\sigma_E/E \approx 0.075/\sqrt{E \;[\GeV]} \oplus 2.5 \%$.
In total, the SpaCal has two interaction lengths which provide additional 
measurements for hadrons.

The calorimeters are 
surrounded by a superconducting solenoid providing a uniform
magnetic field of $1.15$ {\rm T} parallel to the beam axis 
for momentum measurement of charged particles. These are measured
in two concentric jet drift chamber 
modules (CJC), covering the polar angle range  
$ 15^\circ < \theta < 165^\circ$~\cite{cjc}.
A backward drift chamber (BDC) 
aids identification of positrons scattered into the SpaCal calorimeter.

The luminosity is measured using the elastic Bethe-Heitler
process $e p \to e p \gamma$. 
The final state positron
and photon are detected in calorimeters situated close to the
beam pipe at distances of $33$ {\rm m} and $103$ {\rm m} from the interaction
point in the positron beam direction.

\section{Simulation of Standard DIS and $I$-Processes}
Detailed simulation of the H1 detector response to hadronic final
states have been performed for QCD models of the 
standard DIS processes and for QCD $I$-induced scattering 
processes. 

\subsection{Simulation of Standard DIS}

The RAPGAP Monte Carlo~\cite{rapgap} incorporates the
${\cal O} (\alpha_s)$ QCD matrix element and 
models higher order parton emissions
 to all orders in $\alpha_s$ using the concept of parton showers~\cite{shower} 
based on the leading logarithm DGLAP equations~\cite{dglap}, where
QCD radiation can occur before and after the hard subprocess. 
The formation of hadrons is performed using the LUND string model~\cite{lund} 
implemented in JETSET~\cite{jetset}. 
This QCD Monte Carlo is called ``MEPS''.

An alternative treatment of the perturbative phase is implemented
in ARIADNE~\cite{ariadne}, where
gluon emissions are simulated using the colour dipole model (CDM)~\cite{cdm}
by assuming a chain of independently radiating
dipoles spanned by colour connected partons. 
The first emission in the cascade is modified to reproduce the matrix element to first 
order in $\alpha_s$~\cite{ariadneme}. 
The hadronisation is performed using JETSET.
This QCD Monte Carlo is called ``CDM'' in the following.

The CDM and the MEPS Monte Carlo simulations are both interfaced to
the program HERACLES~\cite{heracles} to include
${\cal O}(\alpha)$ electroweak corrections to the lepton vertex, where
$\alpha$ is the electromagnetic coupling.

The Monte Carlo DIS samples have been generated using the
CTEQ4~\cite{cteq4} parton density functions and have been reweighted using a
parametrisation extracted from the recent  H1 measurement 
of the proton structure function \cite{h1f297}.

The MEPS and the CDM models have been comprehensively compared to a variety of
hadronic final state data, and an attempt has been made to optimise the
free model parameters \cite{tuning96,inst:mcws99tuning}.
No single parameter set was found which describes all studied
distributions well. 
Moreover, the Monte Carlo models make different predictions.
Therefore and in view of the involved approximations it is
questionable to what extent the currently available QCD models
can describe 
the standard DIS hadronic final state in particular
in the tails of distributions.
In this analysis the Monte Carlo models have been used with their default 
parameter values. 

\subsection{Simulation of QCD Instanton-induced Processes} 
QCDINS~\cite{inst:qcdins,inst:schremppdis95} is a 
Monte Carlo package to simulate 
QCD $I$-induced scattering processes in DIS.
It acts as a hard process generator embedded in the HERWIG~\cite{herwig}
program.
The hard process is treated according to the physics assumptions
explained in section \ref{sec:qcdtheo}.
Apart from the $Q^2$ cut,
the default parameters of the QCDINS 2.0 version were used, i.e.
$\xprime > 0.35$, $\qprimesq > 113$ \GeVsq and the number
of flavours is set to $n_f=3$.
The CTEQ4~\cite{cteq4} parton density functions have been employed.
After assembling the hard  $I$-subprocess, further QCD emissions
are simulated in the leading-logarithm approximation. The coherent
branching algorithm implemented in HERWIG is used.
The transition from partons to the observable hadrons is performed
with the cluster fragmentation model \cite{inst:cluster}.
%

%
The hadronic final state topology is mainly influenced by the energy
available for the hard $I$-subprocess. 
%
It has been explicitly
checked that the conclusions drawn from this analysis are unchanged
when the LUND string model is used instead of the cluster fragmentation model. 
This has also been observed in Ref.~\cite{inst:mcws99}, 
where in addition the effect of changing free model parameters in the hadronisation 
models has been studied.
For what follows
it is assumed that the commonly used hadronisation models 
are also applicable to describe the fragmentation of a large
number of ${\cal O}(10)$  partons produced by the I-process in a narrow
pseudo-rapidity region with high transverse energy.

\section{Event Selection and  Search Strategy}
\subsection{Inclusive DIS Event Selection}
\label{sec:incldis}
The data used in this analysis were collected in the years
$1996$ and $1997$ with the H1 detector at the electron proton collider HERA.
During this time HERA collided positrons at an energy
of $E_e = 27.5$ \GeV with protons at an energy of 
$E_p = 820$ \GeVx.
The accumulated data sample corresponds
to an integrated luminosity of $21.1$~{\rm pb}$^{-1}$.

The scattered positron is identified as the electromagnetic energy deposition with the highest energy.
For this pioneering analysis we restrict the electron energy and angle measurement to the SpaCal calorimeter.
The electron is required to lie well within the calorimeter and trigger acceptance
of polar angles between $156^\circ$ and $176^\circ$.
A minimal positron energy of $E^\prime_e \ge 10$\GeV is required.
The events are triggered by demanding a localised
energy deposition in the SpaCal together with loose track requirements in the
multi-wire proportional and the drift chambers.

Furthermore  the longitudinal momentum  balance  is required to lie
within $35\,\GeV  < \sum (E - p_z) < 70\,\GeVx$,
where the sum runs over the scattered electron and
all objects belonging to the hadronic final state.
The hadronic final state objects are reconstructed from the
calorimetric energy depositions in the LAr and the SpaCal
calorimeters  and from low momentum tracks ($0.15 < p_t < 2$~\GeVx) 
in the central jet chamber according to the procedure described in \cite{h1highq2}.
The position of the  $z$ coordinate of the reconstructed 
event vertex must be 
within $\pm 30\,\mbox{\rm cm}$ of the nominal interaction point.

The photon virtuality $Q^2$ and the Bjorken scaling variable $x$
are reconstructed from the scattered positron. 
The events are selected to cover the phase space region defined by
$\theta_{e} > 156^\circ$, $ 0.1 < y < 0.6$, $x >10^{-3}$
and $10 \lsim Q^2< 100~{\rm GeV}^2$, where
$\theta_{e}$ is the polar angle of the scattered positron.

The selected DIS data sample consists of about $375000$ events.
The simulated events reproduce well the shape and the absolute
normalisation of the distributions of the 
energy and angle of the scattered positron 
as well as the kinematic variables $x$, $Q^2$ and $y$.
The contamination with events due to hadrons misidentified as positrons
produced in collisions of  high energetic protons with quasi-real photons 
is below $2 \%$. 
This was estimated using the Monte Carlo simulation program 
PHOJET~\cite{phojet}, which
contains the ${\cal O}(\alpha_s)$ matrix elements for 
direct and resolved photon processes, parton showers
and a phenomenological description of soft interactions.

\subsection{Definition of the Discriminating Observables} 
\label{sec:obs}
The observables used to discriminate the $I$-induced contribution from the
standard DIS process are based on the hadronic
final state. 
Only hadronic final state objects as defined in section \ref{sec:incldis}
within  $-1.4 < \eta_{\rm lab} < 2$ are considered.  
Charged particles with transverse momenta of $p_t > 0.15$ \GeV
are selected within $20^o < \theta < 155^o$.
Here, both $\eta_{\rm lab}$ and $ p_t$ are measured in the laboratory frame.

All hadronic final state objects are boosted to the hadronic
centre-of-mass frame\footnote{The hadronic centre-of-mass frame
is defined by
$\vec{\gamma} + \vec{P} = 0$, where $\vec{\gamma}$ ($\vec{P}$) is the $3$-momentum
of the exchanged photon (proton).}.
Jets are defined by the cone algorithm \cite{pozo,cdfcone} 
with a cone radius of $R = \sqrt{\Delta \eta^2 + \Delta \phi^2} = 0.5$.
The jet with the highest transverse energy (\etjetx)
is used to estimate the $4$-momentum $q^{''}$ of the current quark
(see Fig.~\ref{kin-var}).
\qprimesq can be reconstructed from the particles associated with the current jet 
and the photon $4$-momentum, which is obtained  using the measured 
momentum of the scattered positron.
The cone size was chosen to optimise the resolution of the
reconstruction of \qprimesq (see also Ref. \cite{inst:jgerigk}). 
The \qprimesq resolution is about $20 - 30 \%$.
However, the distribution of the true over the reconstructed value
exhibits large tails, since in about $30\%$ of the cases the wrong jet 
is identified as the current jet.
Due to the limited accuracy of the \qprimesq reconstruction, 
the reconstructed \qprimesq cannot be used to experimentally 
control the ``true'' \qprimesq region of the $I$-processes,
but can nevertheless be exploited to discriminate $I$-processes
from the standard DIS background.
The reconstructed \qprimesq is called \qprimesqrec in what follows. 
%
More information on the \qprimesq reconstruction 
can be found in \cite{inst:mcws99,inst:jgerigk,inst:koblitz}. 
The reconstruction of the variable \xprime is more difficult\footnote{ 
The data used for the present analysis do not incorporate a cut on \xprimex. 
As noted in Refs. \cite{inst:rs-lat,inst:schremppzoom},  this presumably 
does not prevent a qualitative comparison with the
Ringwald-Schrempp predictions in the fiducial \xprime and \qprimesq region
for the following reason. 
The lattice data for the $I\overline{I}$-distance $R$ 
distribution (from which the minimal theoretical \xprimex-cut was
deduced  \cite{inst:rs4,inst:rs-lat,inst:schremppzoom}) exhibit a rapid
suppression of $I$-effects for small $I\overline{I}$-separation, corresponding to
$\xprime < 0.35$. Therefore, $I$-contributions to the data from  this $\xprime$
region outside the validity of $I$-perturbation theory can probably be neglected.
} and 
only possible for $I$-events exhibiting the characteristic 
$I$-topology \cite{inst:koblitz}.

The hadronic final state objects belonging to the current jet are not used 
in the definition of the following observables. 
A band in pseudo-rapidity with a width of $1.1$ units in $\eta$ 
is defined around the centre of gravity 
$\bar{\eta} = \sum E_T \eta /(\sum E_T)$
of the transverse energy ($E_T$) distribution 
of the hadronic final state objects
(see Ref.~\cite{inst:koblitz} for details).
This pseudo-rapidity band is called the $I$-band in the following.
The number of charged particles in the $I$-band  
measured as tracks in the detector is
counted (\nbx) and the total scalar transverse energy 
of all hadronic final state objects in the $I$-band is measured (\etbx). 

All hadronic final state objects in the $I$-band are boosted to an
approximate $I$-rest frame defined by $\vec{q'} + \langle \xi \rangle \vec{P} = 0$,
where $\langle \xi \rangle = 0.076$ is
the average value expected by the QCDINS Monte Carlo simulation 
(see Fig.~\ref{kin-var} for definition).
In this system the sphericity (\sphx) 
is calculated\footnote{The sphericity is defined as 
${\rm SPH} = (3/2) (\lambda_2 + \lambda_3)$ where $\lambda_2$ and $\lambda_3$
are the smallest of the three eigenvalues of the diagonalised sphericity tensor
defined by $S^{\alpha \beta} = (\sum_i p_i^\alpha p_i^\beta) / \sum_i {|p_i|}^2$,
where $\alpha$ and $\beta$ corresponds to the $x, y$ and $z$ components
of the considered particle momenta $p_i$.
}. 
For spherical events \sph is close to $1$,
while for pencil-like events \sph is $0$.
Furthermore, the axes
$\vec{i}_{\rm min}$ and $\vec{i}_{\rm max}$
are found for which in the $I$-rest system 
the summed projections of the $3$-momenta
of all hadronic final state objects in the $I$-band are minimal or
maximal \cite{inst:greenshaw}.  The relative difference between
$E_{in} = {\sum_h |\vec{p}_h \cdot \vec{i}_{\rm max} |}$ and 
$E_{out}= {\sum_h |\vec{p}_h \cdot \vec{i}_{\rm min} |}$ 
is called $\deltabx = (E_{in}-E_{out})/E_{in}$.  
This quantity is a measure of the 
transverse energy weighted azimuthal isotropy of an event.  
For isotropic events  \deltab is
small while for pencil-like events \deltab is large.

Three observables are used
to enhance the fraction of $I$-events in the inclusive data sample:
the charged particle multiplicity in the $I$-band (\nbx),  
the sphericity of all final state objects in the $I$-band calculated
in the approximate $I$-rest frame (\sphx) and the reconstructed \qprimesqrecx.
Three other observables, i.e.
\etjetx, \etb and \deltabx, contain additional information and will
be used for further checks.
%
\subsection{Comparison of Data to Standard QCD Predictions in Inclusive DIS} 
\begin{figure}[hb]
\begin{center} 
\epsfig{figure=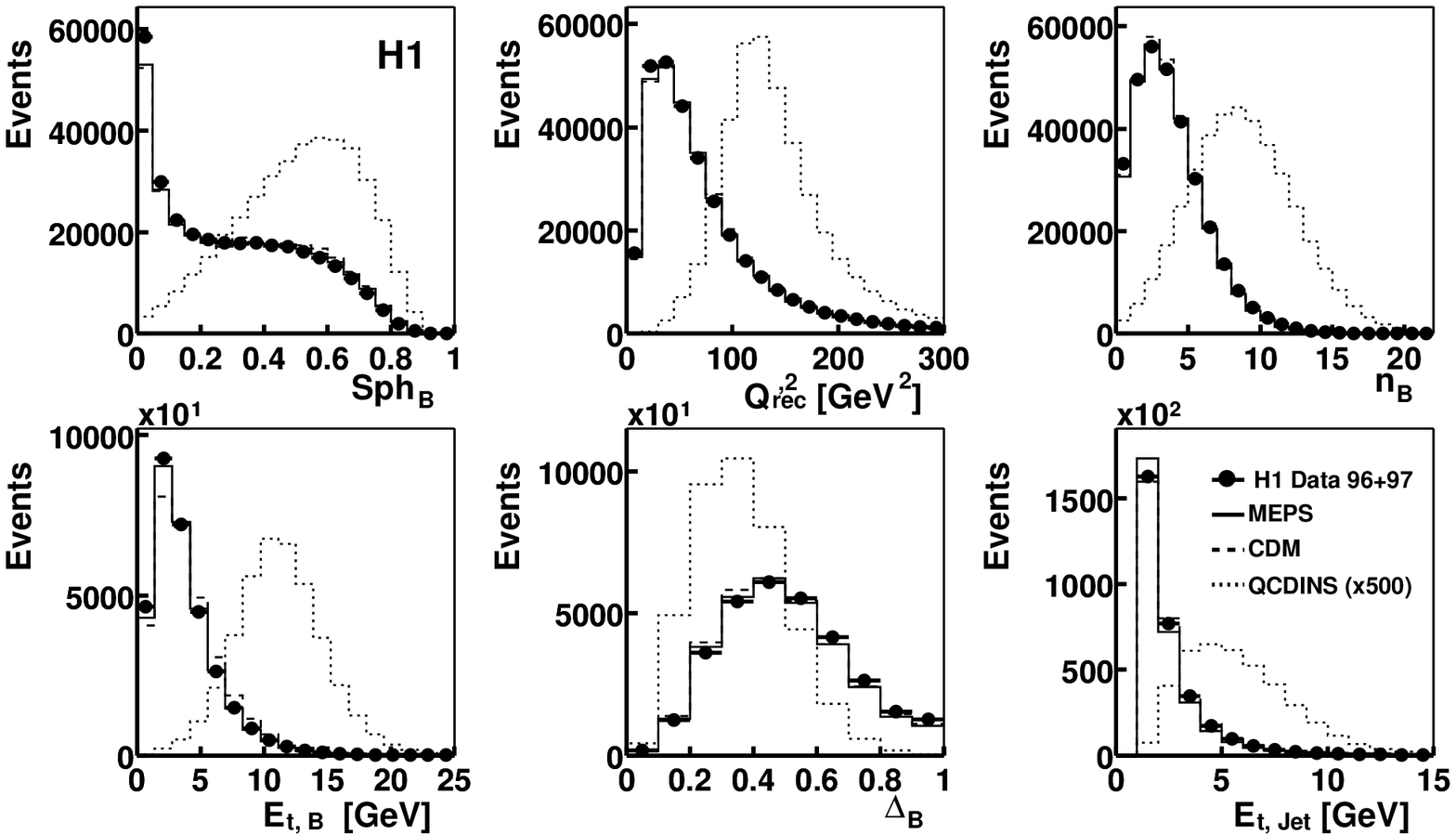,
width=16.cm}
\end{center}
\vspace{-5.3cm}
\begin{picture}(50.,50.)
\put(43., 80.) {a)}
\put(96., 80.) {b)}
\put(150.,80.){c)}
\put(43.,38.) {d)}
\put(96.,38.) {e)}
\put(130.,38.){f)}
\end{picture}
\caption{Distributions of (a) the sphericity in the $I$-band, \sphx, 
(b) the reconstructed virtuality, \qprimesqrecx, 
(c) the charged particle multiplicity in the $I$-band,
(d) the total transverse energy in the $I$-band, \etbx,
(e) the isotropy variable \deltab and 
(f) the transverse current jet energy, \etjetx, in the inclusive DIS sample.
Data (filled circles), the QCD model background Monte
Carlo simulations (solid and dashed line) and the QCDINS
prediction scaled up by a factor of $500$ (dotted) are shown.
\label{fig:datamclin}}
\end{figure}

\begin{figure}
\begin{center}  
\epsfig{figure=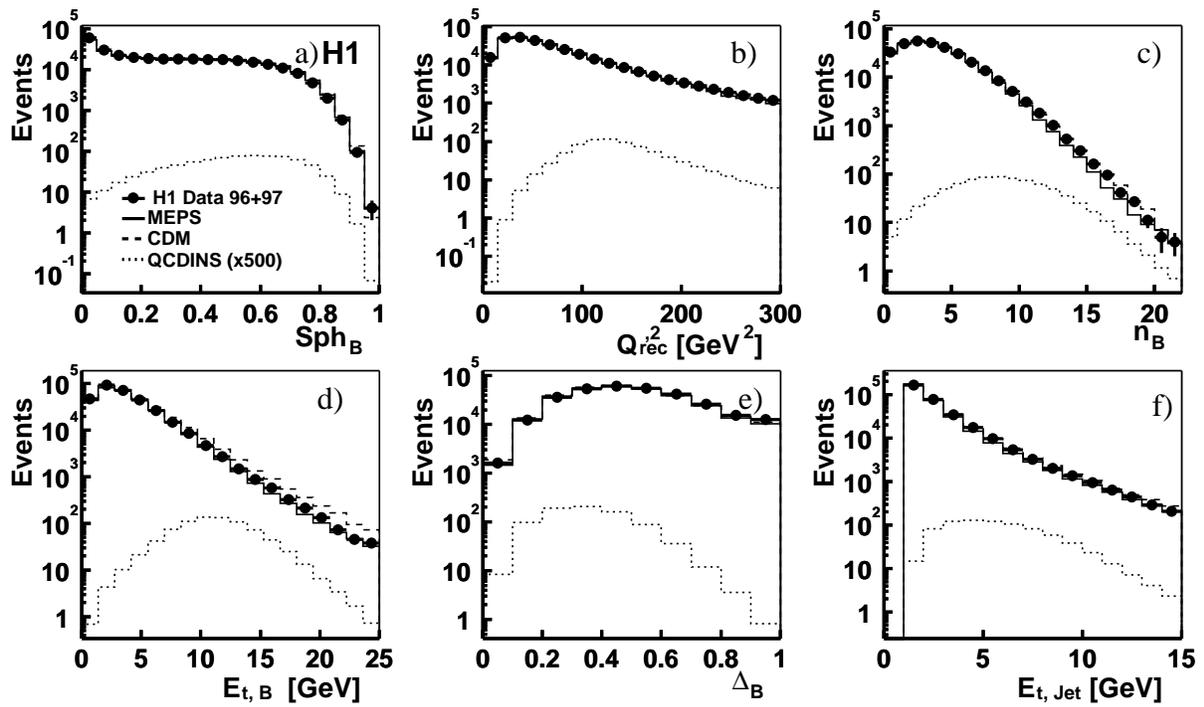,
width=16.cm}
\end{center}
\vspace{-5.3cm}
\begin{picture}(50.,50.)
\put( 40.,87.){a)}
\put( 98.,87.){b)}
\put(152.,87.){c)}
\put( 43.,41.){d)}
\put( 99.,41.){e)}
\put(154.,40.){f)}
\end{picture}
\caption{Same distributions as in Fig.~\ref{fig:datamclin}
except for the absence of rescaling the QCDINS predictions, but on a
logarithmic scale.
\label{fig:datamclog}}
\end{figure}

The distributions of the observables \nbx, \sph and \qprimesqrec
for data, for two standard DIS
QCD models and for the $I$-process are shown 
in Figs.~\ref{fig:datamclin} and \ref{fig:datamclog}.
The gross features of the data are reasonably well described
by both Monte Carlo simulations.  
The CDM model is able to describe the data
within $10\%$ except at very low and very large sphericity
values where a difference of $20\%$ is observed.
The MEPS Monte Carlo  reproduces the data within $10-15$\%.
However, at large \nb deviations up to  $30\%$ are found.

The three other observables, i.e.
\etjetx, \etb and \deltabx, used as control distributions
are shown in Figs.~\ref{fig:datamclin} and \ref{fig:datamclog}.
The \deltab distribution is fairly well described (within $10-20 \%$) 
by both standard DIS simulations.
For \etjet the two standard DIS Monte Carlos simulations 
behave differently.
MEPS describes the data within $10\%$ for $\etjet < 2.5$~\GeVx,
but is lower by about $20\%$ for $\etjet > 2.5$~GeV.
In the tail of this distribution (for $\etjet > 10$\GeVx) 
the data are well reproduced up to the largest accessible values.
The CDM model describes
the data within $5-10 \%$ for $\etjet \lsim 5-10$ \GeVx,
but the predictions progressively grow to be above the data at higher values.
A harder tail than found in the data is also seen in the \etb
distribution. CDM overshoots the data by $50 \%$ at large 
\etb values.
The MEPS simulation gives a much better description
of this observable (within $20 \%$).
The hard transverse energy tail produced by CDM has also been
observed in two-jet and three-jet production 
in DIS at HERA \cite{inst:mcws99tuning,h1jet2,h1jet3}. 

The instanton prediction is shown 
as a dotted line in Fig.~\ref{fig:datamclin} and Fig.~\ref{fig:datamclog}.
For visibility in Fig.~\ref{fig:datamclin} it is scaled up by a factor $500$.
The expected $I$-signal is about two to three
orders of magnitude smaller than the standard DIS background.
Therefore cuts are needed to enhance the signal to background ratio. 
\section{Search for Instanton-Induced Events}
\label{sec:search}
Two methods are employed to increase the sensitivity to $I$-processes: 
a combinatorial cut method described in section~\ref{sec:cuts} and a 
multivariate discrimination technique based on a range 
search method described in section~\ref{sec:disc}.
While the main advantage of the combinatorial cut method is its simplicity,
the multivariate discrimination technique allows the transition from a
background dominated to an $I$-enriched region to be
investigated with one single observable.

\subsection{Combinatorial Cut Based Method}
\label{sec:cuts} 
The strategy to reduce the standard DIS background 
is based on the observables \nbx, \qprimesqrec and \sphx.
These observables have been chosen, since they provide the best 
signal to background separation.
Moreover, in the case of $I$-induced processes the shape of their distributions 
is expected not to be much affected
by contributions from non-planar diagrams \cite{inst:schremppzoom} 
(see also section \ref{sec:qcdtheo}).

To find the optimal combinations of cut values 
the three observables are investigated using simulations 
of the standard DIS background and of the $I$-signal. 
Amongst the  studied cut combinations with an instanton efficiency
$\epsilon_{I} \gsim 10 \%$
the one is chosen with the best separation power, defined by
$S = \epsilon_{I}/\epsilon_{sDIS}$, where 
$\epsilon_{sDIS}$ is the fraction of remaining standard DIS background.

The best  separation power $S$
is obtained for $95 < \qprimesqrec < 200 \GeVsqx$, 
$ \nb > 11$ and $\sph > 0.4$. For this cut combination $S = 125$ ($S = 86$) 
is found for the MEPS (CDM) simulation. 
At an $I$-efficiency  of about $10 \%$,
the background has been suppressed by about a factor of $1000$.
With these cuts, $484$ events are found in the data, while
CDM predicts $443^{+29}_{-35}$ and MEPS $304^{+21}_{-25}$ (see also Tab.~\ref{tab:results}).
The quoted errors on the expected event numbers include 
the statistical and the experimental systematic uncertainties.

The uncertainties arise from the following sources:
a $4 \%$ uncertainty in the hadronic energy scale
of the LAr calorimeter,
$1 \%$ for the electromagnetic energy scale and
$7 \%$  for the hadronic energy scale measured in the SpaCal
calorimeter, $3\%$ for the measurement of the
track momentum, $2$ {\rm mrad} for the polar and azimuthal
angle of the track, $2 \%$ ($5 \%$) for tracks with a momentum
above (below) $250$ {\rm MeV}
for inefficiencies in the track reconstruction 
and $2$ {\rm mrad} for the polar
angle of the scattered positron.
These uncertainties have been propagated into the overall systematic error.
An absolute
normalisation error of $1.5$\% for the accuracy of the luminosity
determination and a $3$\% uncertainty for the reweighting of the
parton densities used in the Monte Carlo simulation
have been included. 
The main contributions to the systematic uncertainties on the number of expected events are seen to arise from the track efficiency ($3.5 \%$) and the momentum measurement ($3\%$).
%
%
\begin{table}[tb]
\begin{center}
\begin{tabular}{|c|l|c|c|c|}
\hline
 \multicolumn{5}{|c|}{Combinatorial cut method (section \ref{sec:cuts} )} \\
\hline  %
        &  N & $\epsilon_{sDIS}$ &   $\epsilon_{I}/\epsilon_{sDIS}$& $\sigma_{lim}$\\
\hline
   Data &  $484$             &            &       &           \\
   CDM  &  $443^{+29}_{-35}$ & $0.118 \%$ &  $86$ & $ 47$ \pb \\ 
   MEPS &  $304^{+21}_{-25}$ & $0.081 \%$ & $125$ & $109$ \pb \\    
\hline
\hline
 \multicolumn{5}{|c|}{Multivariate method (section \ref{sec:disc})}                    \\
\hline  %
        &  N & $\epsilon_{sDIS}$ &   $\epsilon_{I}/\epsilon_{sDIS}$& $\sigma_{lim}$\\
\hline
 Data  & $410$              &            &       &             \\
   CDM & $354^{+40}_{-26}$  & $0.095 \%$ & $106$ & $55$ \pb    \\ 
  MEPS & $299^{+25}_{-38}$  & $0.080 \%$ & $126$ & $80$ \pb    \\     
\hline  
\end{tabular}
\caption{Number of events observed in the data and expected from the CDM
and MEPS simulation after optimising the $I$-signal
to background ratio. 
The quoted error contains the full statistical
and systematic uncertainty added in quadrature. The efficiency
of the standard DIS simulation ($\epsilon_{sDIS}$), the separation
power ($S = \epsilon_{I}/\epsilon_{sDIS}$) and the 
exclusion limit at $95 \%$ confidence level
on the production cross-section of $I$-induced
processes simulated in the fiducial region 
$\xprime > 0.35$ and $\qprimesq > 113 \GeVsqx$ in the
kinematic region defined by
$x > 10^{-3}$, $0.1 < y < 0.6$, $Q^2 < 100$ \GeVsq and $\theta_{e} > 156^o$
are also given.
The $I$-cross-section calculated by $I$-perturbation theory
is $43$\pbx.
\label{tab:results}}
\end{center}
\end{table}

\begin{figure}
\begin{center} 
 \epsfig{figure=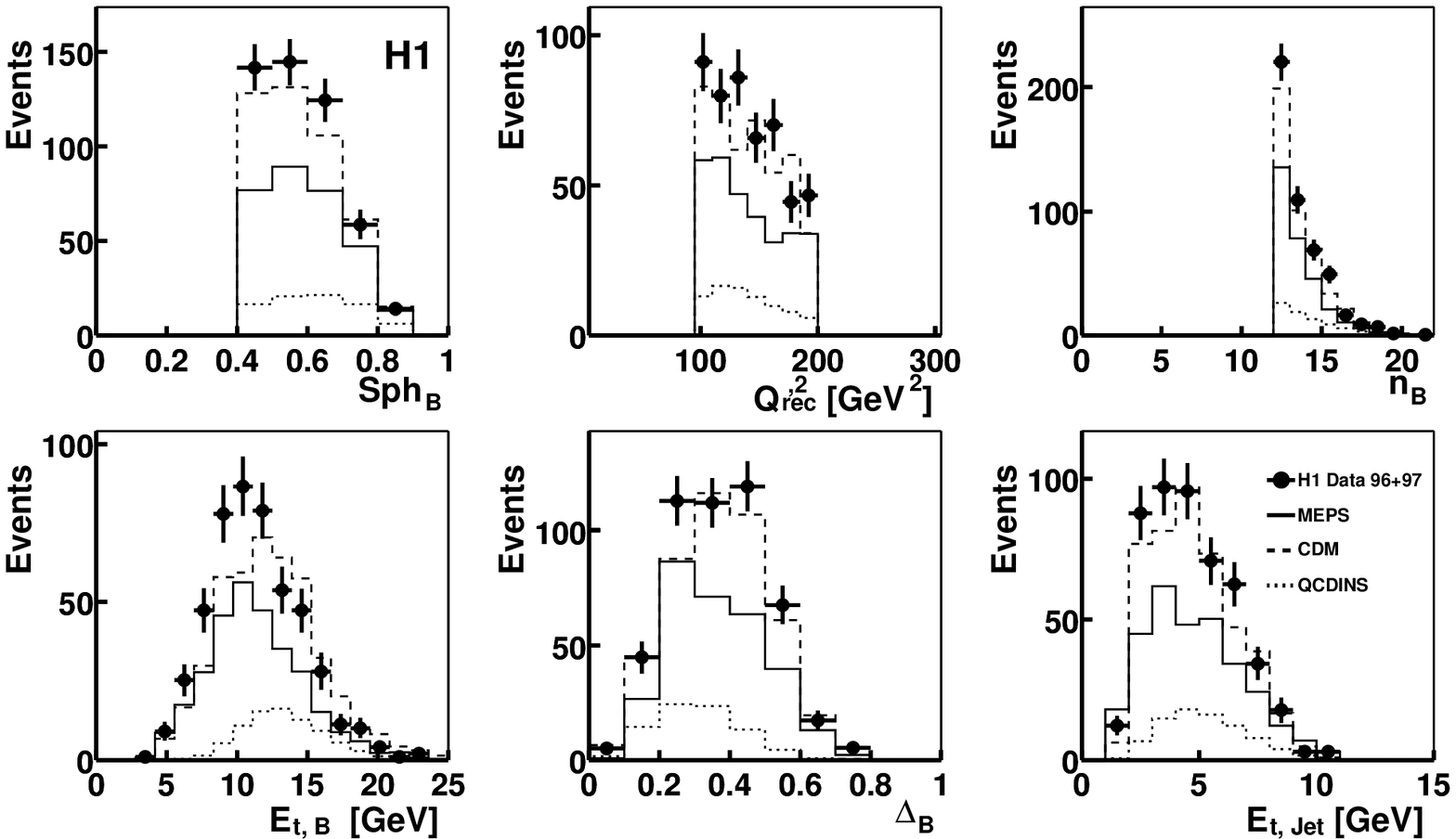,
width=16.cm}
\end{center}
\vspace{-5.3cm}
\begin{picture}(50.,50.)
\put(43. ,80.) {a)}
\put(96. ,80.) {b)}
\put(150.,80.){c)}
\put(43.,38.) {d)}
\put(96.,38.) {e)}
\put(120.,38.){f)}
\end{picture}
\caption{
Distributions of observables after the combinatorial cuts:
(a) the sphericity in the $I$-band, \sphx, 
(b) the reconstructed virtuality of the quark, \qprimesqrecx, 
(c) the charged particle multiplicity in the $I$-band, \nbx, 
(d) the total transverse energy in the $I$-band, \etbx, 
(e) the isotropy variable \deltab and
(f) the transverse current jet energy, \etjetx.
Data (filled circles), the QCD model background Monte
Carlo simulations (solid and dashed line) and the QCDINS
prediction (dotted) are shown.
\label{fig:datamccuts}}
\end{figure}

More events are found in the data than expected by either one of the
background Monte Carlo simulations.
MEPS suggests a clear excess in the data.
In contrast to the findings before placing the cuts
CDM is compatible with the data within errors.
The predictions of the two standard DIS Monte Carlo simulations
largely disagree with each other. This 
indicates that the background estimation 
is subject to large uncertainties as it is expected
in this extreme region of phase space. 
The distribution of the discriminating observables after the cuts
is shown in Fig.~\ref{fig:datamccuts} in comparison with the
standard DIS background expectation from the MEPS and the CDM Monte
Carlo models. Fig.~\ref{fig:datamccuts} also shows the 
expectation for the $I$-contribution to the DIS process as modelled
by QCDINS. 
In particular when compared to the MEPS model,
the shape of the observed excess in the data is 
qualitatively compatible with an $I$-signal for 
\nbx, \qprimesqrec and \sphx.
However, it tends to lie towards low \etb values in contrast to the $I$-contribution. 
It has been noted in~\cite{inst:schremppzoom} that the \etb (as the \etjetx) distribution 
expected for the $I$-signal are most sensitive to theoretical uncertainties.
A shift towards lower values is 
well possible within these uncertainties.
\subsection{Search Based on a Multivariate Discrimination Technique}
\label{sec:disc} 
To make optimal use of the information contained in the
observables separating the $I$-signal and the background a 
multivariate discrimination technique is used.
Events are classified as signal or background by estimating their
probability density $\rho$ at each point in the
phase space of the observables. 
Monte Carlo simulations are employed to sample these densities. 
The densities at each phase space point can be directly estimated by counting the number of
expected signal and background events in a surrounding box. 
The likelihood of an event to be due to a signal can be defined by:
$$D=\frac{\rho(I)}{\rho(I)+\rho(sDIS)}.$$
By cutting on this discriminator $D$,
the background contamination can be minimised at the expense 
of signal efficiency. For each value of $D$ 
the signal to background ratio as well as the signal efficiency
can be easily calculated.

To estimate the $\rho(I)$ and  $\rho(sDIS)$  in the vicinity
of a given event in the multi-dimensional phase space, 
a range search algorithm \cite{sedgewick} based on a binary tree
has been used. 
A more detailed description and the properties of this method can be
found in \cite{BT1,birger}.

\begin{figure}
\begin{center}  
 \epsfig{figure=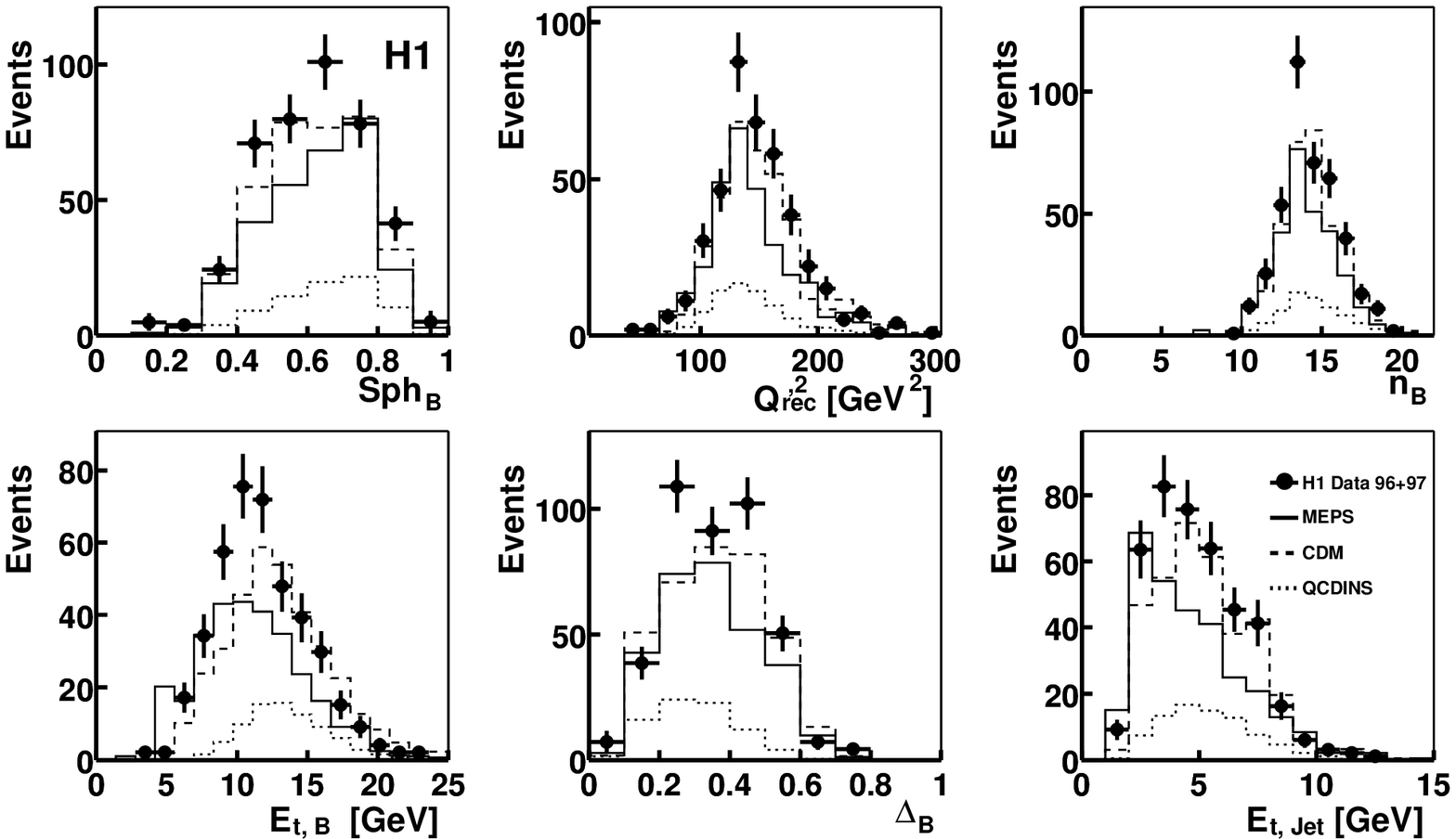,
width=16.cm}
\end{center}
\vspace{-5.3cm}
\begin{picture}(50.,50.)
\put(43., 82.) {a)}
\put(96., 82.) {b)}
\put(150.,82.){c)}
\put(43. ,38.) {d)}
\put(96. ,38.) {e)}
\put(120.,38.){f)}
\end{picture}
\caption{
Distributions of observables after 
a cut on the discriminator $D > 0.988$ to enrich $I$-events:
(a) the sphericity in the $I$-band, \sphx, 
(b) the reconstructed virtuality of the quark, \qprimesqrecx, 
(c) the charged particle multiplicity in the $I$-band, \nbx, 
(d) the total transverse energy in the $I$-band, \etbx, 
(e) the isotropy variable \deltab and
(f) the transverse current jet energy, \etjetx.
Data (filled circles), the QCD model background Monte
Carlo simulations (solid and dashed line) and the QCDINS
prediction (dotted) are shown.
Data (filled circles), the QCD model background Monte
Carlo simulation (solid and dashed line) and the QCDINS
prediction (dotted) are shown.
\label{fig:datamcdisc}}
\end{figure}

\begin{figure}
\begin{center} 
 \epsfig{figure=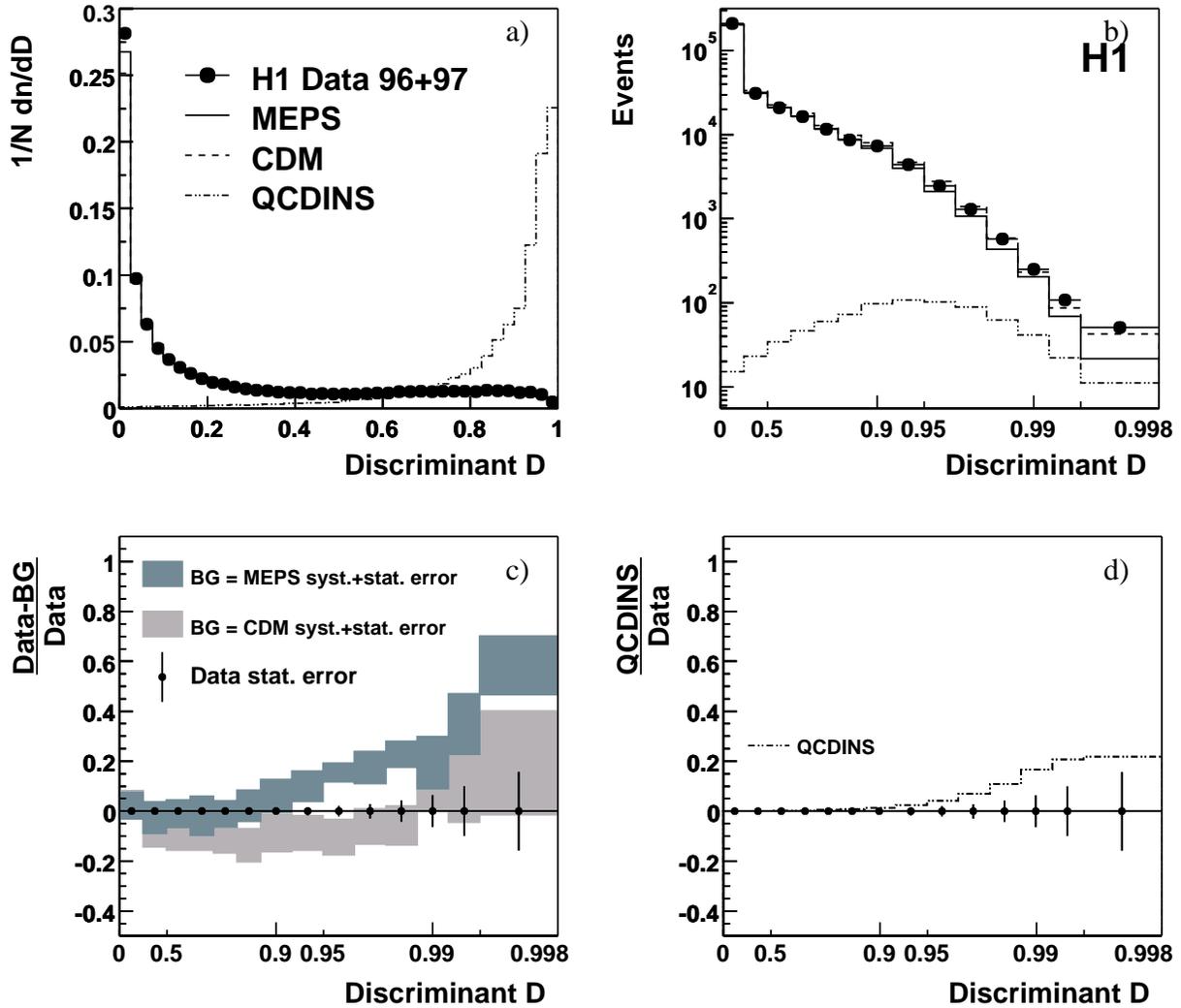,width=16.cm}
\end{center}
\vspace{-5.3cm}
\begin{picture}(50.,50.)
\put( 68.,134.) {a)}
\put(150.,134.){b)}
\put(68. ,60.) {c)}
\put(150.,60.) {d)}
\end{picture}
\caption{(a) normalised and (b) event number 
distribution of the discriminator $D$ and (c) the 
relative difference of data and standard DIS background Monte Carlo simulation.
In (d) the ratio of the QCDINS prediction to the data is shown. 
The bands indicate the experimental uncertainty of the 
standard DIS  background Monte Carlo. The vertical lines in (c) illustrate
the statistical error of the data. In (b), (c) and (d) the $x-$axis is
scaled as $- \log_{10}{(1 -D)}$.
\label{fig:discr}}
\end{figure}

For the discriminator the phase space is spanned by the
same observables \nbx, \qprimesqrec and \sph,
which are used for the combinatorial cut method.
The discrimination power of the method is demonstrated
in Fig.~\ref{fig:discr}a where the shape of the discriminator 
distribution is shown. The simulated background events
(solid and dashed line)
are mainly concentrated at low $D$ values while the simulated $I$-signal events
(dotted line) are peaked towards $D = 1$.
For $\epsilon_{I} = 10 \%$, which corresponds to a cut
at $D > 0.988$, a separation power of
$S = 126$ for MEPS and $S = 106$ for CDM is obtained, respectively.
The separation power is slightly improved with respect
to the combinatorial cut method. $410$ events are observed in the
data, while $354^{+40}_{-26}$ ($299^{+25}_{-38}$) are expected for CDM (MEPS)
(see also Tab.~\ref{tab:results}). These results are consistent with the
ones obtained from the cut-based method.
The dominant contributions to the systematic uncertainties are attributed to the track efficiency ($3.5 \%$), the momentum measurement ($3-6\%$) and the energy scales in the calorimeters ($4-5\%$).
The discriminating observables after a cut $D > 0.988$ are shown
in Fig.~\ref{fig:datamcdisc}. 
Note that the phase space region selected by the discriminator
roughly coincides with the results of the combinatorial cut method.
The observable distributions are suppressed at approximately 
the same points where  
the cuts have been placed by the combinatorial cut method.
The shape of the excess in the data 
in the \nbx, \qprimesqrec and \sph 
distributions is similar to the shape of the expected $I$-distributions.
The excess in the \etb distribution is largest at $\etb \approx 9 \GeV$
while from $I$-processes a peak at about $\etb \approx 12 \GeV$
is expected. 
Again, it should be noted that these distributions might be shifted
towards lower values when contributions from 
non-planar diagrams are taken into account \cite{inst:schremppzoom}. 
The fact that for the \etb and the \etjet distributions 
the standard DIS Monte Carlo simulation
also disagree with each other indicates
large uncertainties from the different treatment of
higher order QCD processes.

The multivariate discrimination technique offers furthermore
the possibility to compare the description of the data by
the Monte Carlo simulation in the complete phase space, i.e.
from the region where no $I$-contribution is expected ($D = 0$)
to the $I$-enriched region ($D = 1$).
%
%
Fig.~\ref{fig:discr}b shows the absolutely normalised
discriminator distribution on a logarithmic $x$-axis 
scaled as $- \log_{10}{(1-D)}$. 
The majority of the standard DIS background
events are concentrated at the lowest $D$ values. 
Towards larger $D$ values
the background falls by three orders of magnitudes.
The data roughly follow this trend. In the last three bins, 
a slight excess of data over background is observed.
According to the QCDINS simulation, in this region
$10$\% of all $I$-events are contained.
As shown in Fig.~\ref{fig:discr}d, where the expected $I$-signal with respect
to the data is shown, the $I$-contribution in the event sample is about $20$\%.

The description of the data by the background Monte Carlo
simulations is illustrated in more detail in 
Fig.~\ref{fig:discr}c, where the relative difference 
of background and data is shown.
The MEPS Monte Carlo gives an excellent description of the
data for $D < 0.90$ which comprises the vast majority of events. 
Towards larger $D$ values an increasingly large excess of events is seen in the data. 
The largest discrepancy of $60 \%$ is found at the largest $D$ value.
The excess of data as a function of $D$ is qualitatively similar to the 
increasing ratio of QCDINS events to the data (see Fig.~\ref{fig:discr}d).
%

From a purely statistical point of view the excess in particular in case of the 
MEPS background estimation is significant. 
However, the uncertainties in the background estimation are largely unknown in this
extreme phase space region.
This is also reflected in the different behaviour of the
CDM Monte Carlo simulation. While it agrees with the data in the 
pure background region $D < 0.2$, it
is not able to describe the data at larger values ($0.25 < D < 0.9$),
where no significant $I$-contribution is expected.
In this region CDM in contrast to MEPS predicts more events than found in the
data.  
It is interesting that both CDM and MEPS 
fall below the data when the separation, i.e. $D$, is largest.
In the case of CDM this observation is, however, not significant
given the experimental uncertainties.
Whether the excess can be explained by 
$I$-processes or whether it is simply due to a deficiency in the description
of the relevant standard DIS process
remains an open question.
Altogether, despite some excess of events in the $I$-signal region, 
the uncertainties in the background 
estimation are too large to draw firm conclusions.

\section{Exclusion Limits for Instanton-induced Processes}
\label{sec:limits} 
Since no significant excess can be claimed upper limits
on the QCD instanton cross-section are derived.
 
The hadronic final state of $I$-induced events
is strongly influenced by the centre-of-mass energy squared, 
$W_I^2 = \qprimesq (1-\xprime)/\xprimex$, 
available for the partons emerging from the $I$-subprocess.
The distributions of the final state topology therefore
crucially depend on both the minimal cut values, above which
$I$-perturbation theory is expected to be valid,
and on the assumed \xprime and \qprimesq distributions,
which are motivated by the validity of $I$-perturbation theory
and which are only under theoretical control for large enough
\xprime and \qprimesqx.

In this section first upper cross-section limits for instantons produced
in the fiducial region $\xprime > 0.35$ and $\qprimesq > 113$ \GeVsq
are derived.
It is assumed that
the \xprime and \qprimesq distributions are correctly described
by $I$-perturbation theory.
This approach is based on the results described in section \ref{sec:search}. 
To be less dependent on theoretical assumptions instantons are independently simulated 
in several bins of approximately constant \xprime and \qprimesq
and the analysis is repeated.
In this way the uncertainty on the assumed shape of
the \xprime and \qprimesq distributions
is minimised and the analysis is extended towards lower \xprime and \qprimesq
value. Here, only the reasonable assumption is made that the topology 
of an $I$-event is for a given $W_I^2$
correctly modelled by the $I$-Monte Carlo simulation.

\subsection{Exclusion Limits in the $I$-Fiducial Region}
\label{sec:limitsfiducial} 
First an upper limit on the $I$-cross-section 
is derived at $95\%$ confidence level (CL) for instanton produced 
in the fiducial region,
where the \xprime and \qprimesq distributions
are calculated within the Ringwald-Schrempp approach. 
The number of observed events in the data and 
           of expected standard DIS background events 
and the $I$-signal efficiency 
obtained with the combinatorial cut method are used.
To derive the cross-section limits 
the method described in Ref.~\cite{limits} was used.
For the background estimation and the selection efficiency
statistical and systematic errors are taken into account
by folding Gaussian distributions into the integration
of the Poisson law used to determine the limit.

An $I$-cross-section of $109$ \pb ($47$ \pbx)
is excluded when MEPS (CDM) is assumed to provide the correct
background description. These results together with the ones obtained
from the multivariate discrimination technique are summarised
in Tab.~\ref{tab:results}. Both methods lead to similar results.
The limits are not far from the predicted $I$-cross-section of about $43 \pbx$.

As said before, it is questionable whether the CDM and MEPS 
models are able to adequately describe the standard DIS 
background in this extreme corner of phase space, 
where only $\sim 0.1 \%$ of the events in the total sample of standard DIS events
are expected. 
To be independent of the detailed modelling of the hadronic
final state of DIS events, an additional upper limit is extracted
where the expected standard DIS background is assumed to be zero.
Whatever the ``true'' number of standard DIS events in
the selected corner of phase space is, the $I$-signal
can certainly not be bigger than the number of observed
events in the data. An upper limit on the $I$-cross-section derived in this way
is therefore the most conservative one,
since it only uses the expected topology of $I$-induced events and
the number of events observed in the data.
Here the combinatorial cut method was used. 
At $95\%$ confidence level a cross-section of $221$\pb
is excluded without relying on the correct modelling of the 
background. 
The $I$-cross-section predicted within the Ringwald-Schrempp framework
is about a factor of $5$ lower.

\subsection{$I$-Model Independent Exclusion Limits}
\label{sec:limitsnomodel} 
To minimise the theoretical input in the extraction of an
upper limit on the $I$-cross-section,
small ranges of approximately constant \qprimesq and \xprime
are analysed. 
Events in a $5 \; \times \; 5$ grid
with  $0.2 \le \xprime \le 0.45$ (grid size $0.05$) and
with $60 \le \qprimesq \le 160$ \GeVsq (grid size $20$ \GeVsqx) 
have been simulated using the QCDINS Monte Carlo program.
In this way the effect of the assumed shape 
of the \xprime and \qprimesq distributions is minimised and
the analysis is extended into regions, where
the \xprime and \qprimesq distributions cannot be calculated.
It is only assumed that the hadronic final state for instantons
produced at fixed \qprimesq and \xprimex, i.e. at fixed $W_I^2$,
is correctly modelled by the $I$-Monte Carlo simulation.

\begin{figure}
\begin{center} 
 \epsfig{figure=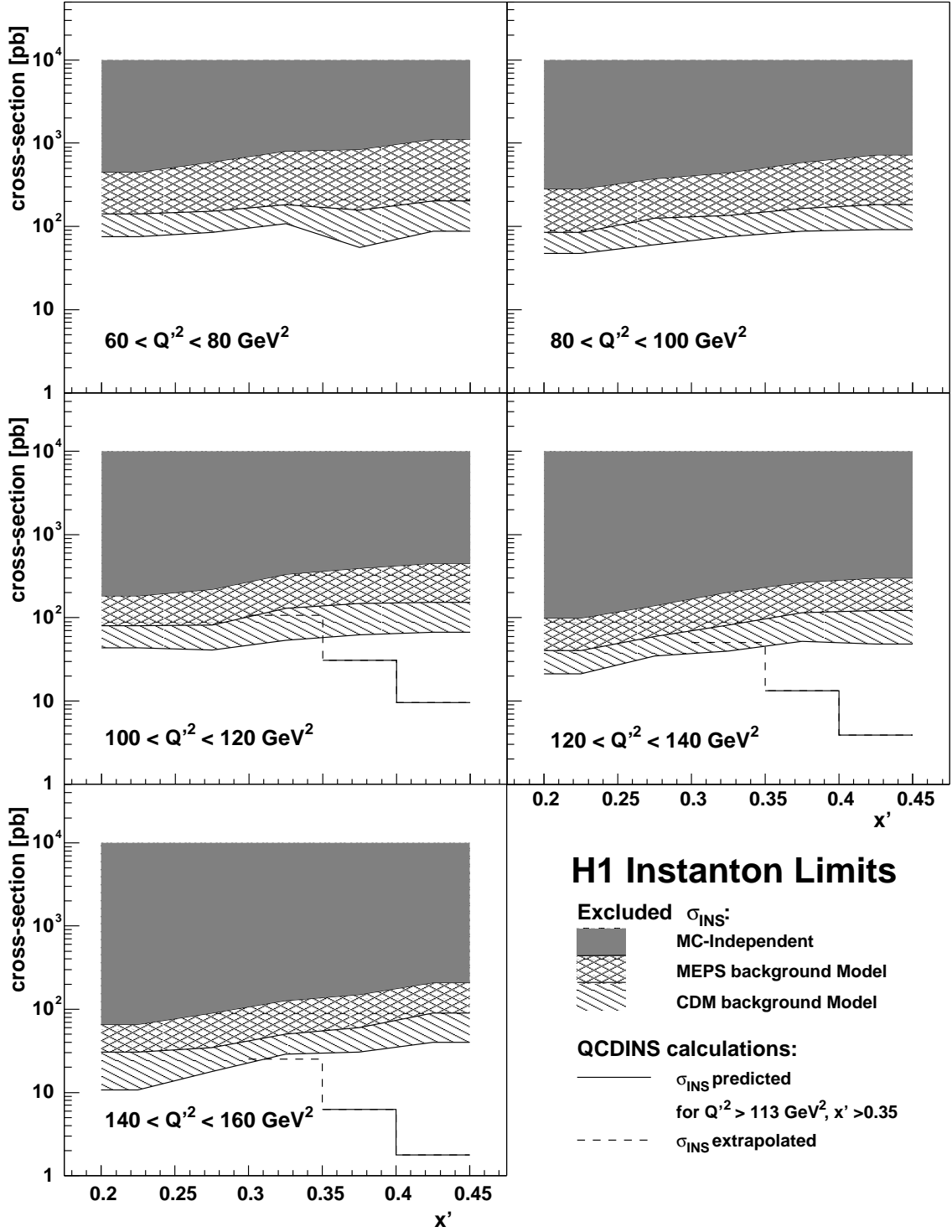,width=15.cm}
\end{center}
\caption{Upper limit on the cross-section for instanton-induced
events as modelled by QCDINS for $I$-events produced in bins 
of \xprime and \qprimesq. 
Regions above the curves are excluded at $95 \%$ confidence level.
The instanton cross-section calculated in the
fiducial region $\xprime > 0.35$ and $\qprimesq > 113$ \GeVsq
(solid line) is also shown
and continued towards lower \xprime (dashed line). 
The limit is valid in the kinematical region of
$x >10^{-3}$, $0.1< y <0.6$ and $\theta_{e} >156^\circ$.
\label{fig:slim}}
\end{figure}

For each particular \xprime and \qprimesq bin, the analysis described 
in section \ref{sec:cuts} is repeated.
The new best cut combinations ensuring an $I$-efficiency
$\epsilon_{I} > 10 \%$ and the maximal separation power
$\epsilon_{I}/\epsilon_{sDIS}$ are chosen. The data selected
for these optimised cuts on \nbx, \qprimesqrec and \sph
are compared to the standard DIS background simulations.
For each case an upper limit on the $I$-cross-section at $95\%$ CL is derived.
In Fig.~\ref{fig:slim} the cross-section exclusion limit
obtained with the CDM and MEPS background simulations 
are shown for each \xprime and \qprimesq region. 
The difference of the results for the two standard DIS models
reflects the uncertainty in the DIS prediction.
The conservative limit independent of the background description, i.e. for the background
assumed to be zero, is also shown. 
Depending on the \xprime and \qprimesq intervals
$I$-cross-sections between $60$\pb and $1000$\pb are excluded. 

In Fig.~\ref{fig:slim} also the $I$-cross-section
as evaluated in the Ringwald-Schrempp framework is shown 
in the region where $I$-perturbation theory is expected to be valid,
i.e. for large \xprime and \qprimesqx corresponding to small instanton sizes.
The upper limits are above the predicted $I$-cross-sections in the fiducial region 
$\xprime > 0.35$ and $\qprimesq > 113$ \GeVsqx. 
The continuation of the fast increase of the
$I$-cross-section towards lower values of \xprime
and \qprimesq as expected from an extrapolation 
of $I$-perturbation theory is also indicated. 
%

\begin{figure}
\begin{center}
 \epsfig{figure=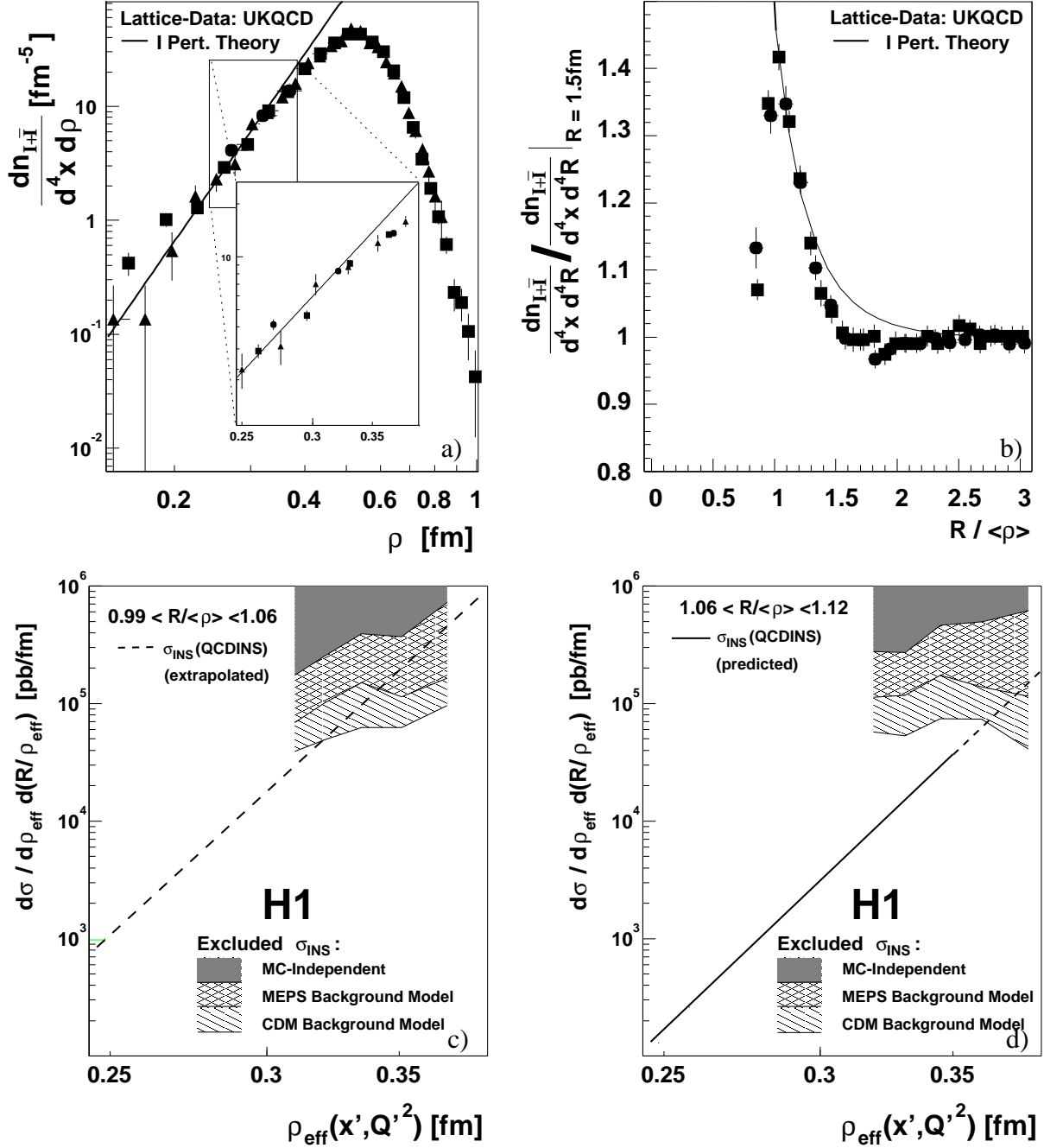,,width=16.cm}
\end{center}
\vspace{-5.3cm}
\begin{picture}(50.,50.)
\put( 67.,108.){a)}
\put(153.,108.){b)}
\put(68. ,17.) {c)}
\put(154.,17.) {d)}
\end{picture}
\caption{Instanton density distribution on the QCD vacuum
as a function of the instanton size $\rho$ (a) and (c) (zoomed)
and of the instanton-anti-instanton distance $R$ distribution 
as a function of \rrho 
normalised to the value at $R = 1.5$~{\rm fm} (b).
Lattice data by the UKQCD collaboration \cite{inst:rs-lat,UKQCD}
are shown as closed symbols,
the prediction of perturbative instanton theory as lines \cite{inst:rs-lat}. 
The dashed vertical lines shows the lower edge of the region
where the deviation of the perturbative calculation and of the
lattice simulations set in.
In (d) the instanton cross-section in two bins of \rrho 
as the function of the instanton size $\rho$ is shown.
The shaded area indicate the experimental limits which is based
on the MEPS background description.
The figures a and b are adapted from Ref.~\cite{inst:rs-lat}.
\label{fig:latticelim}}
\end{figure}

The prediction of $I$-perturbation theory can be compared
with non-perturbative lattice simulations of the QCD vacuum 
for zero flavours\cite{inst:rs-lat,UKQCD} to obtain independent 
information on the $\rho$ and the \rrho ~distributions.
This is illustrated\footnote{The notation is as follows:
$\frac{d n_I}{d^4x \; d\rho}$ corresponds to $D(\rho)$ and
$\frac{d n_{I\overline{I}}}{d^4 x d^4 R} = 
\int_0^{\infty} d\rho \int_0^{\infty} d\bar{\rho} D(\rho) D(\bar{\rho}) \;
 e^{-\frac{4\pi}{\alpha_s} \Omega}$ in eq.~\ref{eq:IXsec}. The variable
$x$ denotes the Euclidian space-time coordinates.
}
in Fig.~\ref{fig:latticelim}a and Fig.~\ref{fig:latticelim}b.
For low $\rho$ and large \rrho,
where the $I$-perturbation theory is expected to be valid, 
a power-like rise $ \propto \rho^6$ (see equation (\ref{eq:pIDensity})) is found
which is (including the absolute normalisation)
in excellent agreement with the non-perturbative lattice simulations.
For $\rho \, \gsim \, 0.35~{\rm fm}$ and $\rrho \, \lsim \, 1.05$ the $I$-perturbative
calculation continues to rise while the lattice data flatten and finally
drop towards $\rho \to 1~{\rm fm}$ and small \rrho, respectively.
In this region the $I$-perturbation theory is not reliable anymore.
Figure~\ref{fig:latticelim}c shows an expanded version of the region where
$I$-perturbation theory starts to deviate from the lattice data.

To compare this observation with the HERA cross-section limits 
we transform the \xprime and \qprimesq bins
into the $\rho$ and \rrho ~bins by 
using the effective $I$-size $\rho_{\rm eff}(\xprimex,\qprimesqx)$,
which dominates the integration in equation (\ref{eq:IXsec})
and can be obtained from information accessible in QCDINS.
To present the cross-section limits we choose one \rrho ~bin which is just
outside ($0.99 < \rrho < 1.06$)
and one bin which is just inside ($1.06 < \rrho < 1.12$)
the fiducial region. $0.99 < \rrho < 1.06$ corresponds to 
$0.3 < \xprime < 0.35$ and $1.06 < \rrho < 1.12$ corresponds to
$0.35 < \xprime < 0.4$.
The obtained cross-section limits are shown in
two bins of \rrho ~in Fig.~\ref{fig:latticelim}d as a function of $\rho_{\rm eff}$.
The QCDINS predictions are shown as lines. 
For $1.06 < \rrho <1.12$, i.e. within the fiducial region of
$I$-perturbation theory, the cross-section limits exclude the continuation
of the power-like rise of the $I$-cross-section with $\rho$ for
$\rho > 0.35 \,{\rm fm}$. 
For $0.99 < \rrho <1.06$ the rise has to be attenuated even earlier.
This disfavours a continuation of a steep rise of the $I$-cross-section
towards large $\rho$ values.
The absence of this rise is in accord with the expectation of lattice
simulation of the QCD vacuum.

\section{Summary}
QCD instanton-induced processes as
modelled in a Monte Carlo simulation implementing the prediction
of $I$-perturbation theory
have been searched for
in deep-inelastic scattering at HERA in the kinematic range
$x >10^{-3}$, $0.1< y <0.6$, $\theta_{e} >156^\circ$ and $Q^2 < 100$\GeVsqx. 
Three observables were used to enhance the sensitivity to instanton events
with respect to the standard DIS QCD background:
the charged particle multiplicity in the instanton rapidity band, 
the reconstructed quark virtuality \qprimesqrec and 
the sphericity of the hadronic final state objects in the instanton rapidity band.
Applying either cuts or a multivariate discrimination
technique based on these three observables, the standard DIS background is
suppressed by typically a factor of $1000$, 
while $10 \%$ of the $I$-events are expected to be kept.
In this region $484$ events were observed and 
$443^{+29}_{-35}$ and $304^{+21}_{-25}$
were predicted by two standard DIS QCD background models.

Using a multivariate discrimination technique 
the transition from the background dominated region to 
the $I$-enriched region is investigated.
The matrix element plus parton shower model 
describing the data in the background
region clearly falls below the data in the signal region.
With increasing sensitivity to $I$-processes 
an increasingly large
excess is seen in the data. The shape of the excess 
is qualitatively compatible with the expected $I$-signal. 
The colour dipole model does not describe the data well in
the background dominated region, but is in better agreement in the
signal region. 
Although the data exceed the expectations where the sensitivity to the instanton
process is expected to be largest,
this effect is not significant given the uncertainties in the background estimation.

The standard DIS Monte Carlo models 
have known deficiencies and fail to describe various aspects of
DIS data in the HERA regime.
A better understanding of the formation of the hadronic
final state in DIS in the bulk and in the extreme end of the 
phase space relevant for instanton searches will be needed
to make further progress.

Based on solely the number of observed events in the data to remain
independent of the modelling of the standard DIS background,
a most conservative upper limit on the instanton cross-section 
of $221$\pb is excluded at $95\%$ confidence level. 
This limit is valid in the fiducial region of instanton perturbation theory,
i.e. for $\xprime > 0.35$ and for $\qprimesq > 113~{\rm GeV}^2$
implying small $I$-sizes $\rho$ and large $I\overline{I}$ distances $R$, 
where the $I$-calculations are expected to be valid.
The limit is about a factor of five above the cross-section predicted by 
instanton perturbation theory
and relies on the $I$-topology based on
the calculated \xprime and \qprimesq distributions.
To be independent of this  assumption
additional upper exclusion limits on the $I$-cross-section have been derived
in fixed \xprime and \qprimesq intervals corresponding to small
regions of $\rho$ and \rrho.
These limits are only based on the
hadronic final state $I$-topology for a given \xprime and \qprimesq
and on the number of observed events in the data.
Depending on the considered kinematic region
$I$-cross-sections between $60$ and $1000$~\pb
are excluded at $95 \%$ confidence limit. 
The limits cannot exclude the predicted $I$-cross-section
in the fiducial region, 
but exclude a steep $I$-cross-section rise with decreasing \xprimex, i.e.
towards large $I$-sizes, as would be obtained from a naive 
application of $I$-perturbation theory in this region. 
The absence of such a steep rise is in accord
with lattice simulations for zero flavours of the QCD vacuum. 

In summary, this initial experimental study has shown that very large instanton contributions at small momentum transfers can be excluded at HERA.
To reach sensitivity at the level of the predicted instanton-induced cross-section requires improved sophistication in the experimental methods and a thorough understanding of the hadronisation process.

\section*{Acknowledgements}
We thank A. Ringwald and F. Schrempp 
for many helpful discussions and for their support with
the QCDINS Monte Carlo simulation program.
We are grateful to the HERA machine group whose outstanding efforts
have made and continue to make this experiment possible.
We thank the engineers and technicians for their work in constructing and now 
maintaining the H1 detector, our funding agencies for financial support,
the DESY technical stuff for continual assistance, and the DESY directorate
for the hospitality which they extend to the non-DESY members of the
collaboration.



%
%

%

%

%

\end{document}